# Effect of Ca doping on Li ion conductivity of Ge and Ta doped garnet LLZO

**Muktai Aote[a], A. V. Deshpande[a,*]**

[a] Department of Physics, Visvesvaraya National Institute of Technology, South Ambazari Road, Nagpur, Maharashtra - 440010, India

[*] **Corresponding Author:**

**Dr. (Mrs.) A. V. Deshpande**

Professor and Head,

Department of Physics,

Visvesvaraya National Institute of Technology,

South Ambazari Road, Nagpur,

Maharashtra, 440010 (India)

E-mail Address: avdeshpande@phy.vnit.ac.in

Ph.No.: +91-712-280-1251

**Abstract**

The series $Li_{6.55+x}Ge_{0.05}La_{3-x}Ca_xZr_{1.75}Ta_{0.25}O_{12}$ ( x= 0, 0.05, 0.10, 0.15, 0.20 ) was prepared by conventional solid state reaction method with the sintering temperature of $1050^0C$ for 7.30 hr by substituting Ca at the La site to increase the Li ion conductivity. The synergistic effects of Ca incorporation on $Li_{6.55}Ge_{0.05}La_3Zr_{1.75}Ta_{0.25}O_{12}$ were studied using various structural and electrochemical analyses. X-ray diffraction, scanning electron microscopy, and Impedance analysis were used to determine the crystal structure, morphology, and AC conductivity of the prepared ceramic samples, respectively. The highest conductivity of 9.95 x $10^{-4}$ S/cm was obtained for a 0.05 Ca ceramic sample with minimum activation energy of 0.23 eV. The DC polarization measurements confirmed the dominance of ionic conduction in 0.05 Ca ceramic. The results obtained make the 0.05 Ca ceramic sample a promising candidate as solid electrolytes for all solid state Li-ion batteries (ASSLIBs).

## 1. Introduction

Electrochemical devices are needed for energy conversion and storage to utilize the abundance of renewable energy sources in nature. Li-ion batteries have gained widespread recognition due to their extensive use in consumer electronic appliances and the electric vehicle industry [1–3]. Li-ion batteries offer high specific capacity and extended duty cycles. However, the organic liquid electrolytes that are being used in these batteries are facing many safety challenges. Specifically, the Li-ion battery used in electric vehicles is susceptible to flammability issues resulting from the dendrite growth in conventional liquid electrolytes, which can lead to short-circuiting [4, 5]. To address this concern, researchers are actively exploring substituting these liquid electrolytes with solid-state electrolytes to develop safer Li-ion batteries. To replace the liquid electrolytes, solid electrolytes must possess specific properties like high room temperature ionic conductivity, a wide electrochemical potential window, a comparable specific capacity, a prolonged duty cycle, and stability with lithium metal anode [6].

A comprehensive exploration of various solid electrolytes, including polymer, inorganic, sulfide, and thin film solid electrolytes, has been conducted. Moreover, solid electrolytes like $Li_{0.35}La_{0.55}TiO_3$ (LTT), $Li_{1.4}Al_{0.4}Ti_{1.6}$ $(PO_4)_3$ (LATP), $Li_7La_3Zr_2O_{12}$ (LLZO), $Li_{1.5}Al_{0.5}Ge_{1.5}$ $(PO_4)_3$ (LAGP) and superionic sulfide like $Li_2S$-$P_2S_5$ are thoroughly investigated. However, all these electrolytes, except garnet structured LLZO, were constrained in achieving the required conductivity range or compatibility with lithium metal [6]. Oxide-type garnet structured $Li_7La_3Zr_2O_{12}$ exhibits an ionic conductivity in the range of $10^{-4}$ S/cm at room temperature and exceptional chemical stability when in contact with lithium metal with a wide potential range (> 5 V/Li) [7]. Nevertheless, the LLZO must exhibit the cubic phase (Ia-3d) to obtain the mentioned conductivity range. Initially, LLZO has the tetragonal phase at room temperature ($I4_1$/acd) with lower ionic conductivity. After suitable heat treatment, this phase transforms into the conducting cubic phase. Although the cubic LLZO improves the conductivity range, it is still less than the conventional liquid electrolytes.

Various strategies have been adapted to enhance Li ion conductivity in LLZO. These approaches include the incorporation of external cations in the lattice of LLZO, the addition of a sintering additive to minimize the temperature range required for stabilizing the cubic phase, and the introduction of glass ceramics in the LLZO to increase the density, which tends to promote conductivity [8–10]. The addition of cation within the lattice of LLZO can occur at any of the sites among Li, La, and Zr. Initially, significant research efforts were focused on the insertion of single supervalent cations such as Al, Ga, Ge, Ba, Mn, Be, Zn, Fe, and W. This approach created the Li-ion vacancies, which eventually led to enhanced ionic conductivity [6, 11–18]. After the successful improvement in conductivity through single-cation doping, the co-doping strategy has been explored. Li, La, and Zr sites have been substituted simultaneously with the external supervalent cations. This study revealed that the synergistic effect is created on the Li-ion conduction in garnet LLZO due to the various co-doping combinations like Ta and Ge, Y and Ga, Sr and Nb, Rb and Ca, Rb and Ta, Al and Ga, Ga and Y [6],[19–23]. Following this trend, researchers are curious about the impact of tri-doping these supervalent cations in the oxide type garnet LLZO. Tri doping involves the substitution at Li, La, and Zr sites simultaneously with three supervalent cations. Although limited studies have been reported with the tri-doping methodology, the outcomes are highly promising for ASSBs [24, 25]. Thus, it is interesting to investigate and explore the effect of such doping on the structural and electrochemical properties of the LLZO.

In the present study, the effect of Ge, Ta, and Ca substitution in LLZO has been investigated. The supervalent cations Ge and Ta have been selected based on the results of the earlier reported studies [6]. The primary objective of adding Ge and Ta was to obtain the conductive cubic phase at a relatively lower temperature than previously reported studies. Ge has proven its utility as a sintering aid. At the same time, Ta plays an essential role in stabilizing the cubic phase [26]. Here, Ca has been selected to substitute at the La site. The substitution of alkaline earth metal elements (Ca, Sr, Ba, Mg) at the La site in LLZO can compensate for the decreased Li content caused by the insertion of supervalent cations like Ge and Ta in garnet LLZO. It increases the total Li content, which helps to enhance the overall Li ion conductivity [3, 27]. Among the various alkaline earth elements, Ca gained special attention due to the minimal difference in its ionic radius (1.12 $A^0$) compared to Lanthanum (1.16 $A^0$). It is expected to minimize the possible lattice distortion in LLZO and promote Li-ion conductivity. Also, it was observed that the optimum content of Ca can help in the densification of the ceramic, which is an essential property for the ease of ionic conductivity [3]. Table 1 show the maximum ionic conductivity obtained with the doping of $Ca^{2+}$ in garnet LLZO.

In this work, $Li_{6.55}Ge_{0.05}La_3Zr_{1.75}Ta_{0.25}O_{12}$ composition with high Li ion conductivity is further investigated with the insertion of $Ca^{2+}$ at the La site. The fixed concentration of Ge and Ta helps maintain the optimum Li concentration of 6.4-6.6 atoms per formula unit, as it is essential to enhance the ionic conductivity in LLZO [6]. The Ca content has been varied between 0 – 0.20 and the corresponding synergistic effect on structural, chemical, and electrochemical properties has been studied.

**Table 1: Maximum ionic conductivity reported with the doping of $Ca^{2+}$ in garnet LLZO.**

| Composition | Method | Sintering Time & Temperature | Ionic conductivity S/cm | References |
|---|---|---|---|---|
| $Li_{6.65}La_{2.95}Ca_{0.05}Zr_{1.6}Sb_{0.4}O_{12}$ | Solid state reaction | $1175^0$C/24 h | $8.95 \times 10^{-4}$ | 3 |
| $Li_{6.42}Ca_{0.02}La_{2.98}Zr_{1.4}Ta_{0.6}O_{12}$ | Solid state reaction | $1230^0$C/ 1 h | $5.69 \times 10^{-4}$ | 28 |
| $Li_{6.45}Ca_{0.05}La_{2.95}Zr_{1.4}Ta_{0.6}O_{12}$ | Solution method | $1125^0$C/6 h | $4.03 \times 10^{-4}$ | 29 |
| $Li_{7.1}\ La_3Zr_{1.95}Ca_{0.05}O_{12}$ | Solid state reaction | $1235^0$C/ 16 h | $8.95 \times 10^{-4}$ | 30 |
| $Li_{6.9}La_{2.6}Ca_{0.22}Zr_{1.82}Nb_{0.25}O_{12}$ | Sol-gel method | $800^0$C/ 2 h | $9 \times 10^{-4}$ | 31 |
| $Li_{6.3}La_{2.9}Ca_{0.2}\ Zr_{1.4}Ta_{0.6}O_{12}$ | Sol-gel method | - | $3.5 \times 10^{-4}$ | 32 |

## 2. Experimental Work

### *2.1. Sample Preparation*

The series $Li_{6.55+x}Ge_{0.05}La_{3-x}Ca_xZr_{1.75}Ta_{0.25}O_{12}$ with x varying from 0 to 0.20 was synthesized using the conventional solid state reaction method. All the necessary chemicals $Li_2CO_3$ (Merck, >99.9%), $CaCO_3$ (Sigma Aldrich, >99.0%), $La_2O_3$, $ZrO_2$, $GeO_2$, and $Ta_2O_5$ (Sigma Aldrich, >99.99%) were weighed according to the stoichiometry of the series. The chemicals were then hand mixed in an agate mortar. 10% excess of $Li_2CO_3$ was added to the powder mixture to compensate for the Li loss due to high sintering temperature. For the calcination process, the powder sample was transferred into an alumina crucible and placed in a muffle furnace at $900^0$C for 8 h. After calcination, the powder was again finely crushed. The pellets with dimensions of around 10 mm diameter and 1.5 mm thickness were formed using a hydraulic press with uniaxial pressure of 4 tons. The formed pellets were sintered in the bed of mother powder at $1050^0$C for 7.30 hr with the crucible covered by an alumina lid in an air atmosphere. Each sample in a series is represented according to its Ca content as 0 Ca, 0.05 Ca, 0.10 Ca, 0.15 Ca, and 0.20 Ca.

### *2.2 Sample Characterizations*

The sintered pellets were subjected to various structural and electrochemical characterizations. The phase identification was done by powder X-ray diffraction (RIGAKU diffractometer). The data were collected with a step size of $0.02^0$ in the range of $10^0$- $70^0$ and a scan speed of $2^0$/min. The Cu-kα with the wavelength of 1.54 $A^0$ was used as a radiation source. Archimedes' principle was followed to determine the densities of ceramic samples (K-15 Classic (K-Roy) instrument) using toluene as an immersion medium. The surface morphological study was conducted using the Scanning Electron Microscope (SEM) technique with the JSM-7600 F/JEOL instrument. The electrochemical analysis was done using a Novocontrol impedance analyzer within the frequency range of 20 Hz to 20 MHz. The data collected between the temperature range of $25^0$C to $150^0$C was utilized to calculate the activation energy of ceramic samples. Additionally, the DC polarization study was carried out to determine the ionic transport number with the KEITHLEY 6512 electrometer.

## 3. Results and Discussion

### *3.1. X-ray Diffraction*

The X-ray diffraction patterns for all the synthesized ceramic samples with varying Ca content are shown in Fig.1 (a). It is observed that the conductive cubic phase has been achieved for all the samples. The cubic phase determination was done from the reference JCPDS file no. 45.0109 of garnet structured $Li_5La_3Nb_2O_{12}$. This indicated that the garnet LLZO structure could accommodate supervalent cations of different sizes without affecting its geometrical structure [25]. The earlier reported study mentioned the formation of the $CaZrO_3$ phase for the 0.20 Ca doping content [3]. In contrast, in this current study, such a phase was not observed, resulting in the pure cubic structure ceramic sample without any impurity phase. This can be achieved due to minimal Ta and Ge doping, which helps to stabilize the cubic phase at lower sintering temperature [6], along with the insertion of $Ca^{2+}$ at the $La^{3+}$ site, which introduced more $Li^+$ to get occupied at the Li site which maintains the charge neutrality [33]. Moreover, the addition of $Ca^{2+}$ causes an increment in the sharpness of the peak (422) along with the intensity. It indicates the enhancement in crystallinity with no such lattice deformation. Hence, this result suggests that the optimal concentration of $CaCO_3$ with Ta and Ge does not affect the cubic nature of garnet LLZO. In the current study, Ca has been incorporated at the La site. This successful site incorporation can be seen from Fig.1 (b). The figure shows the magnified image of the XRD patterns from $42^0$ to $44^0$ for all the prepared ceramic samples (Ca = 0 – 0.20). The (611) peak is shifting towards the higher angle with the increasing Ca content, suggesting a decrease in lattice parameters. This can be attributed to the insertion of a lower ionic radius, Ca2+ (1.12 $A^0$), at the site of La3+ (1.16 $A^0$), which has a larger ionic radius.

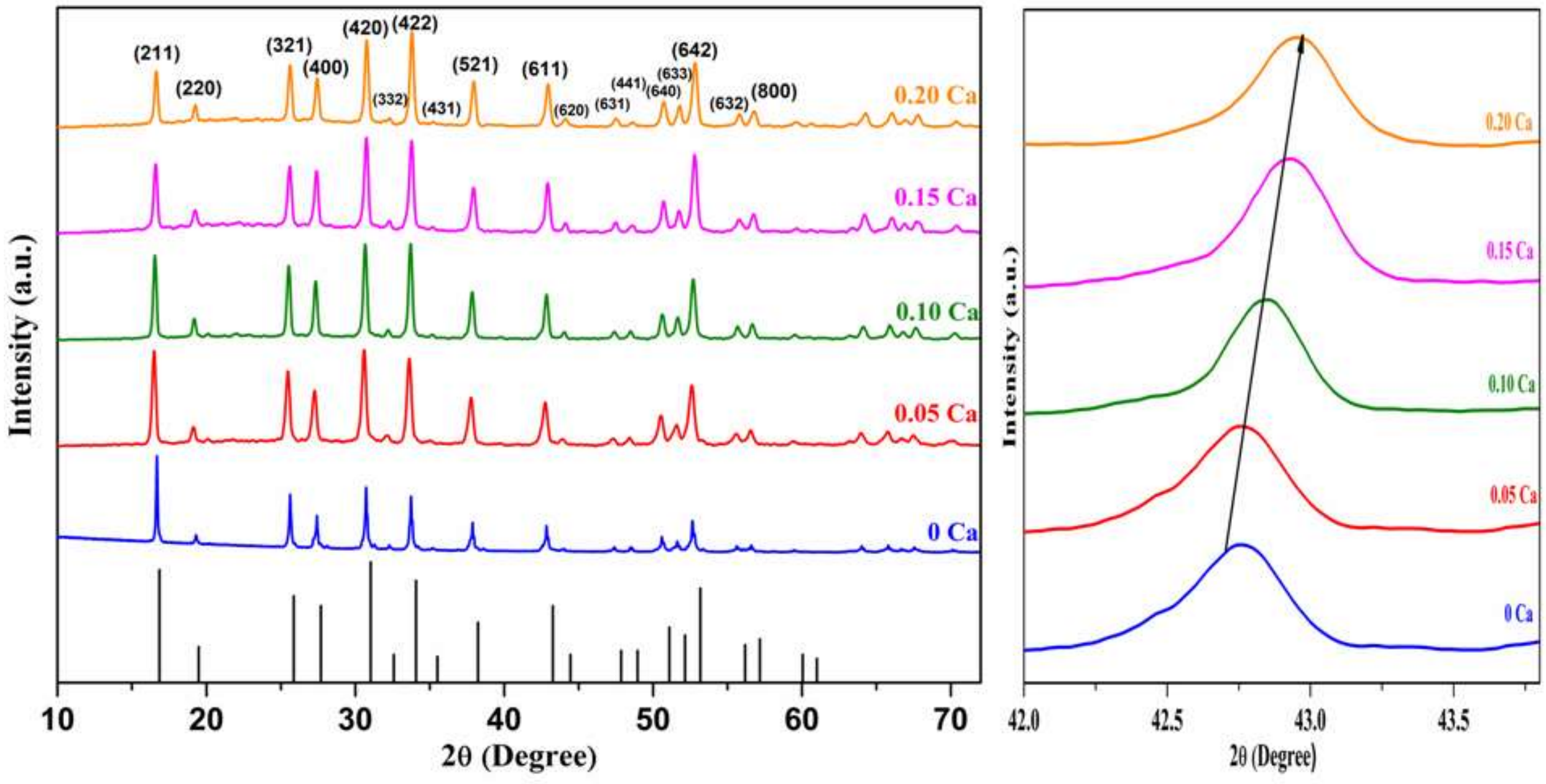

**(a)** **(b)**

**Fig.1: (a) X-ray diffraction pattern of $Li_{6.55+x}Ge_{0.05}La_{3-x}Ca_XZr_{1.75}Ta_{0.25}O_{12}$ (x = 0 – 0.20).**

**(b) Shifting of (611) peak towards higher angles.**

### *3.2. Density Measurement*

The densities of all the synthesized ceramic samples in $Li_{6.55+x}Ge_{0.05}La_{3-x}Ca_xZr_{1.75}Ta_{0.25}O_{12}$ series with varying Ca content from 0 to 0.20 have been measured using Archimedes' principle. Toluene was used as an immersion liquid. Fig. 2 shows the variation in the densities of prepared samples with Ca content. From the figure, it can be observed that the highest density is achieved for the composition having 0.05 Ca. All the other samples also exhibit relative densities greater than 90%. However, a further increase in Ca content leads to a slight decrease in density. The 0.05 Ca ceramic sample possesses the highest density value with a relative density of 95.09%. This result is in good agreement with the earlier reports, which suggested that the optimal Ca content can help in the densification of the ceramic sample[3, 28]. The samples with Ca content exceeding 0.05 shows slightly reduced densities due to the formation of pores within the structure, which can be observed from surface morphology images in Fig.3.

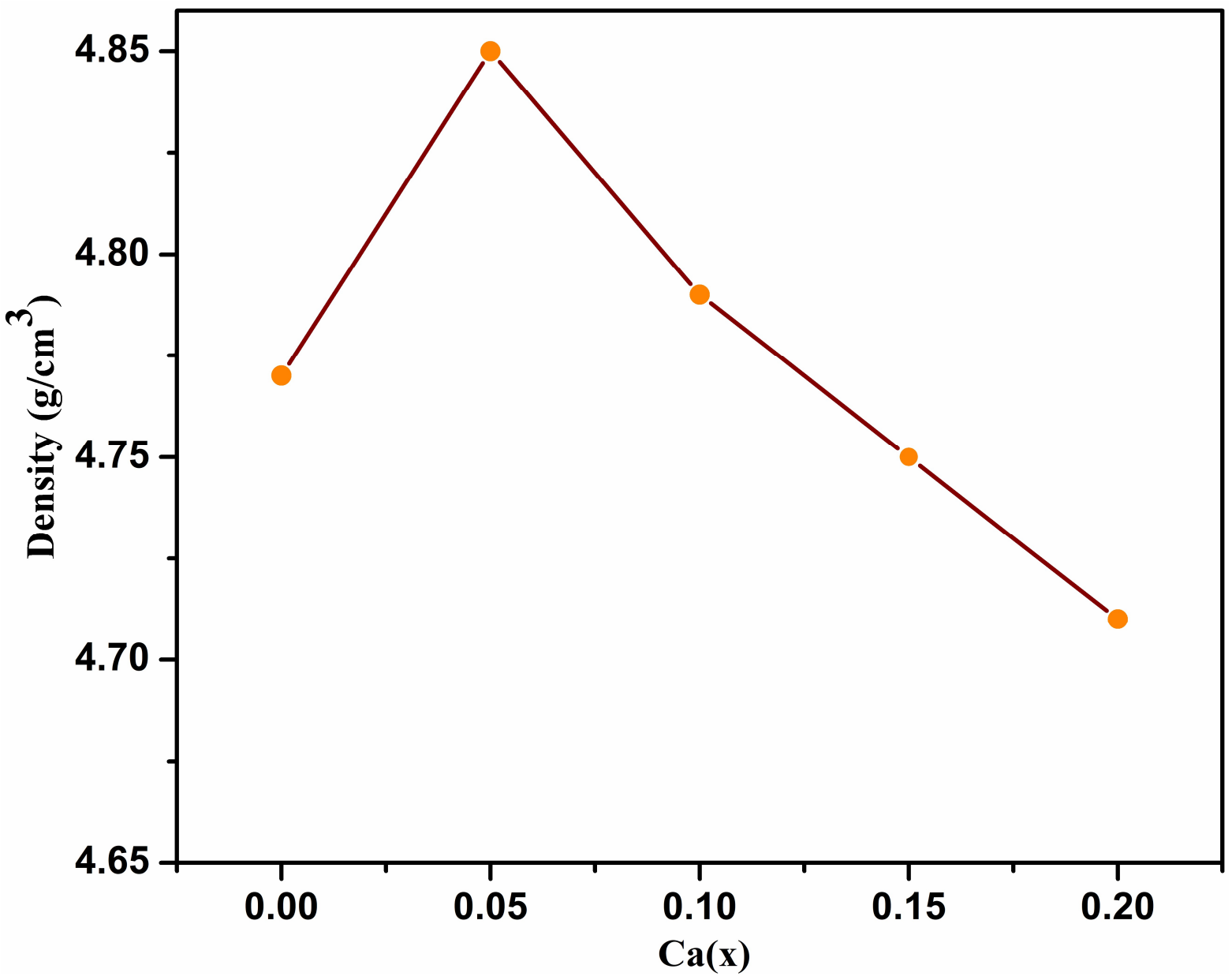

**Fig.2: Variation in density with the varying content of Ca in**

**$Li_{6.55+x}Ge_{0.05}La_{3-x}Ca_xZr_{1.75}Ta_{0.25}O_{12}$**

### *3.3. Surface Morphological Study*

The surface morphology of all the ceramic samples (x= 0 – 0.20) sintered at $1050^0$ C for 7.30 h can be seen in Fig. 3. Fig.3 (a) depicts the micrograph of as synthesized sample with 0 Ca. The pores can be clearly observed, affecting the sample's density. The grains have grown in size with 0.05 content of Ca and are well connected with the neighboring grains. There are no pores observed within the structure. This results in dense and compact microstructure formation, as shown in Fig. 3 (b). This result is in good accordance with the obtained value of density and relative density, which suggests that 0.05 Ca content helps in the densification, giving denser microstructure, ultimately allowing for Li-ion migration [6]. With the increase in Ca content, the grains have grown in size, but voids have also been generated within the structure, which can be seen in Fig. 3 (c –e). It can be correlated with the variation in density, as shown in Fig. 2. Thus, the dense structure obtained with 0.05 Ca content can be attributed to the optimum incorporation of lower valence Ca. The optimal incorporation of Ca played a crucial role in increasing grain growth by accelerating the grain boundary migration with the material transfer within the structure [28, 34]. Fig. 3 (f) shows the average particle size distribution of the 0.05 Ca ceramic sample calculated using Image-J software, which was found to be $4.32 \pm 0.07 \mu m$.

Elemental mapping of the 0.05 Ca ceramic sample is shown in Fig. 4. All the constituent elements, such as La, Zr, Ca, Ta, and Ge, are uniformly distributed over the surface of the sample with no chemical segregation [3]. The incorporation of Ca within the structure is also confirmed by the XRD peak shifting towards a higher angle observed in Fig. 1 (b). This suggests that Ca is successfully incorporated in the garnet structured $Li_{6.55+x}Ge_{0.05}La_{3-x}Ca_xZr_{1.75}Ta_{0.25}O_{12}$.

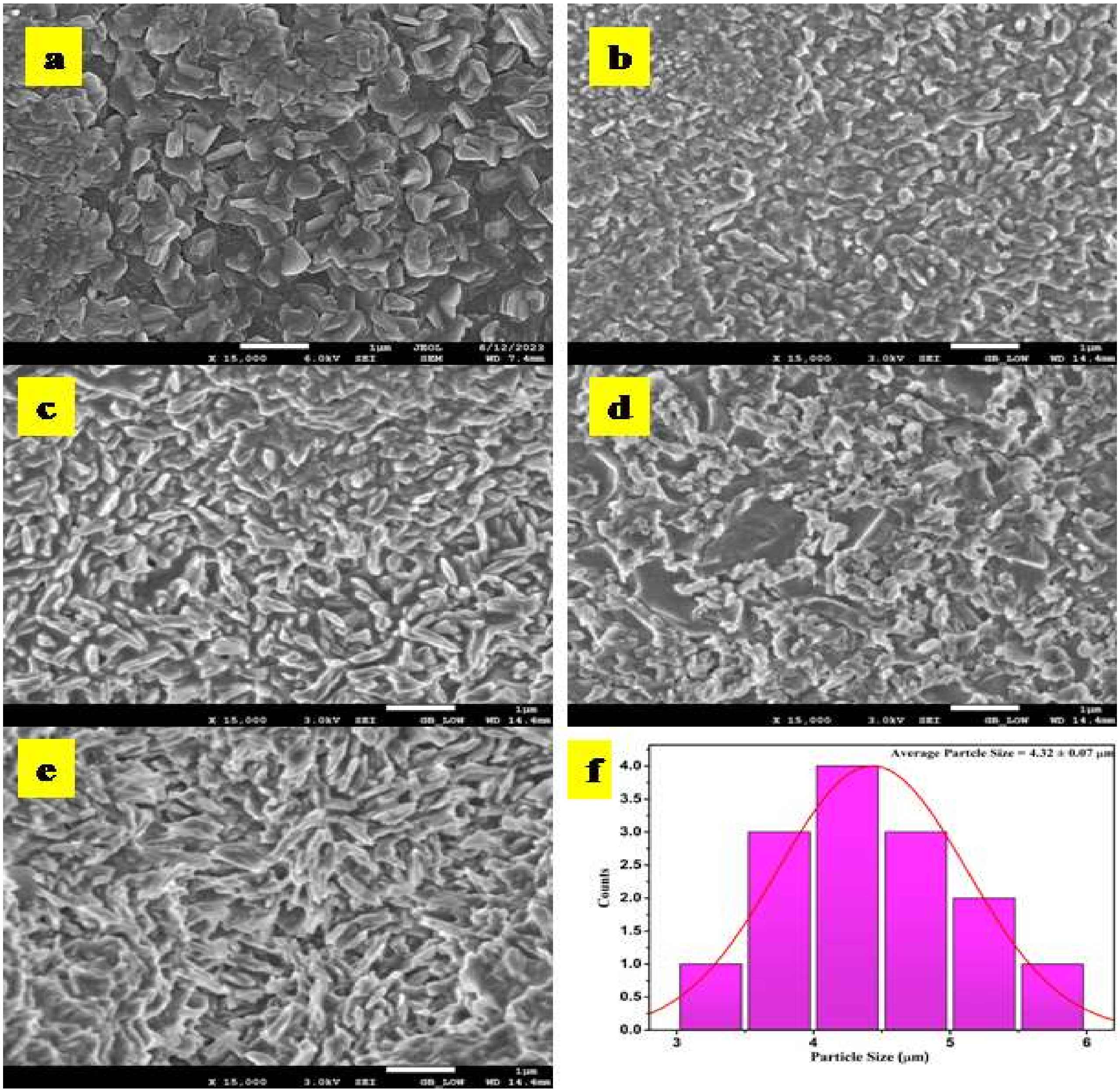

**Fig.3: Surface morphology of $Li_{6.55+x}Ge_{0.05}La_{3-x}Ca_xZr_{1.75}Ta_{0.25}O_{12}$ with (a) x = 0, (b) x = 0.05, (c) x = 0.10, (d) x = 0.15, (e) x= 0.20 and (f) Average particle size distribution of 0.05 Ca ceramic sample.**

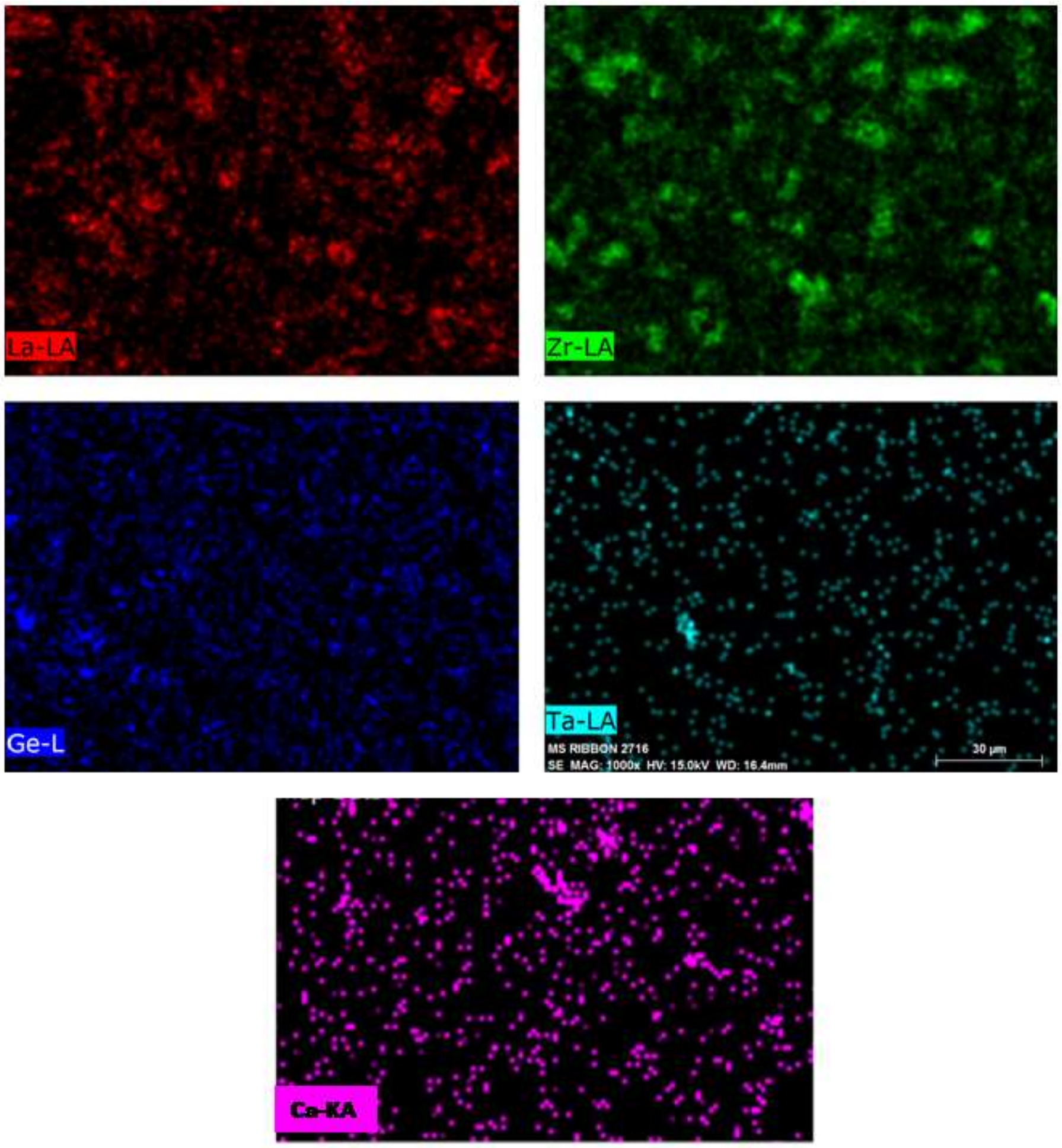

**Fig. 4: Elemental mapping of 0.05 Ca ceramic.**

### *3.4. AC Conductivity Studies*

#### *3.4.1. Impedance Plots*

Fig.5 (a) displays the Nyquist plots of $Li_{6.55+x}Ge_{0.05}La_{3-x}Ca_xZr_{1.75}Ta_{0.25}O_{12}$ series with Ca content ranging from 0 to 0.20 at room temperature. The impedance offered by each ceramic sample can be calculated from the intercept made on the x-axis by the semicircle in the higher frequency region. The obtained semicircle in the higher frequency region is ascribed to the bulk resistance of the sample [31]. The tail can be observed as we move towards the lower frequency. The appearance of the inclined tail in the lower frequency region can be attributed to the nature of the Ag electrodes, which served as ion-blocking electrodes. The existence of ionic conduction is initially confirmed by the obtained nature of the graph [6, 35, 36]. The ionic conductivity offered by all the synthesized ceramics is calculated using the relation as $\sigma_{total} = {}^{t}/_{RA}$, where $t$ is the thickness of the pellet, $R$ is the resistance offered by the sample, $A$ is the area of the electrode, and $\sigma_{total}$ is the ionic conductivity of the sample. Fig. 5 (a) shows that the 0.05 Ca ceramic sample offers the minimum impedance, whereas the semicircle shifts towards a higher impedance value with increased Ca content. The inset graph shows the comparison between 0 Ca and 0.05 Ca. The inset graph depicts that 0.05 Ca offers the minimum resistance with the maximum ionic conductivity of 9.95 x $10^{-4}$ S/cm among the synthesized ceramic samples at room temperature. It can be achieved due to the appropriate amount of Ca incorporation in the garnet LLZO lattice along with Ta and Ge, as it helps in the densification [3]. The result is well supported by the findings of the highest density and dense microstructure obtained for 0.05 Ca, reducing the chances of dendrite growth formation. The connectivity between the grains decreases the intergranular impedance and promotes Li-ion migration. Moreover, incorporating $Ca^{2+}$ at the $La^{3+}$ site increases the total Li content, indirectly enhancing the ionic conductivity by maintaining the optimum Li content. Fig. 5 (b) represents the fitted graph of a 0.05 Ca ceramic sample with the equivalent fitting circuit. $R_1$ and $R_2$ give the grain and grain boundary resistance values, respectively. With the increase in Ca content beyond 0.05 Ca, there is a decrease in ionic conductivity. It can be attributed to the formation of pores in the structure, which can be seen in Fig. 3 (c-e). The presence of voids within the structure also results in lower density. It obstructs the Li-ion conduction pathways, resulting in higher impedance. Also, from this result, it can be suggested that the excessive Li content could not always result in higher ionic conductivity of garnet LLZO [37]. Hence, the discussion confirmed that the optimum content of 0.05 Ca with fixed Ta and Ge content for $Li_{6.55+x}Ge_{0.05}La_{3-x}Ca_xZr_{1.75}Ta_{0.25}O_{12}$ series exhibited the highest ionic conductivity.

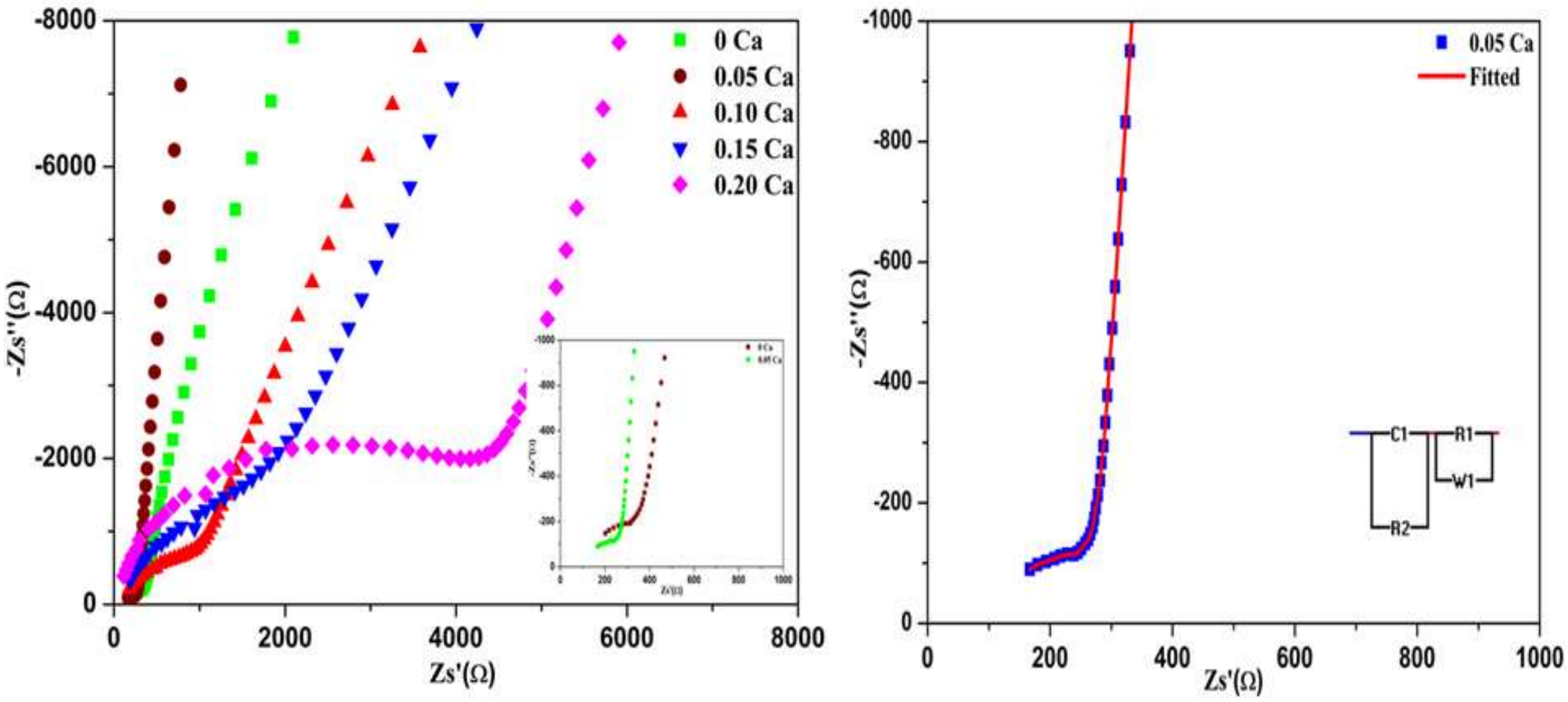

**(a)** **(b)**

**Fig. 5: (a) Nyquist plots of $Li_{6.55+x}Ge_{0.05}La_{3-x}Ca_xZr_{1.75}Ta_{0.25}O_{12}$ (x = 0 – 0.20) at room temperature. (b) Fitted spectra of 0.05 Ca ceramic sample.**

### *3.4.2. Arrhenius Plots*

Following the impedance measurement, fig. 6 (a) represents the Arrhenius plots of all the prepared ceramic samples of $Li_{6.55+x}Ge_{0.05}La_{3-x}Ca_xZr_{1.75}Ta_{0.25}O_{12}$ (x = 0 – 0.20) series. The measurements were done in the temperature range of $50^0$C to $150^0$C. The activation energies were calculated from the Arrhenius equation, i.e., $\sigma(T) = \sigma_0 \exp\left(\frac{-E_a}{K_B T}\right)$, where $\sigma$ is the conductivity, $\sigma_0$ is the pre-exponential factor, $E_a$ is the activation energy, $K_B$ is the Boltzmann constant, and $T$ is the temperature in Kelvin. Fig. 6 (a) shows that the minimum activation energy is obtained for the 0.05 Ca ceramic sample, which also possessed the highest ionic conductivity among all the samples. The minimum activation energy of 0.23 eV is associated with the highest conductivity of 9.95 x $10^{-4}$ S/cm. Activation energy values for all the samples with respective ionic conductivity are mentioned in Table 2. The table shows that, as the Ca content increased from 0.05 to 0.20, the ionic conductivity decreased from 9.95 x $10^{-4}$ S/cm to 5.72 x $10^{-5}$ S/cm, respectively. The highest conductivity result can be attributed to the densification due to 0.05 Ca, which helps Li-ion migration with improved mobility. Also, it can be stated that the optimum content of 0.05 Ca can help in reducing the energy barrier of Li ion. It can be attributed to the valence gap and the difference in the ionic radius of $La^{3+}$ and $Ca^{2+}$ [28]. Fig. 6 (b) shows the variation in activation energy and ionic conductivity with Ca content. Hence, incorporating Ca at the La site played an important role in increasing the ionic conductivity of the Ta and Ge co-doped garnet LLZO.

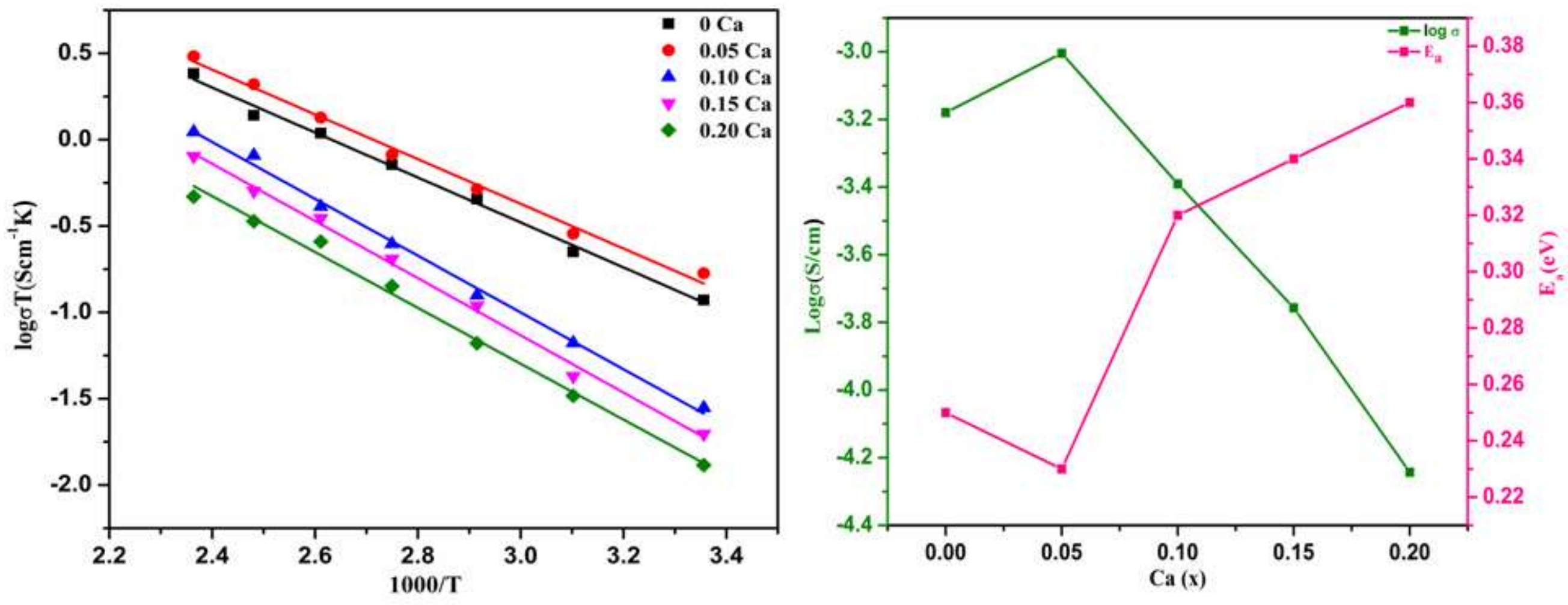

(a) (b)

**Fig.6: (a) Arrhenius plots of $Li_{6.55+x}Ge_{0.05}La_{3-x}Ca_xZr_{1.75}Ta_{0.25}O_{12}$ with x = 0 to 0.20. (b) Variation in ionic conductivity and activation energy with Ca(x) content.**

**Table 2: Values of ionic conductivity and activation energy of $Li_{6.55+x}Ge_{0.05}La_{3-x}Ca_xZr_{1.75}Ta_{0.25}O_{12}$.**

| Content of Ca (x) | Ionic Conductivity (S/cm) | Activation Energy (eV) |
|---|---|---|
| 0 | $6.61 \times 10^{-4}$ | 0.25 |
| **0.05** | $\mathbf{9.95 \times 10^{-4}}$ | **0.23** |
| 0.10 | $4.07 \times 10^{-4}$ | 0.32 |
| 0.15 | $1.75 \times 10^{-4}$ | 0.34 |
| 0.20 | $5.72 \times 10^{-5}$ | 0.36 |

### *3.5. DC Conductivity Study*

DC conductivity was measured with respect to time and current to verify the contribution of electronic conductivity to total conductivity. Fig. 7 (a) shows the DC conductivity plot for the 0.05 Ca ceramic sample. A ceramic pellet was placed between the Ag electrodes to measure electronic conductivity, and a constant voltage of 1 V was applied. The current through the sample was measured with an equal interval of time. After some time, the current through the sample was almost constant, which is only due to electrons [6, 9]. The ionic transport number was calculated using the formula, $t_i = (\sigma_{total} - \sigma_e)/\,\sigma_{total}$ for all the ceramic samples of $Li_{6.55+x}Ge_{0.05}La_{3-x}Ca_xZr_{1.75}Ta_{0.25}O_{12}$ series. The comparison between the ionic and electronic conductivities is shown in Fig. 7 (b). For all the prepared ceramic samples, the ionic transport number was > 0.999, which confirmed that the conduction within the samples is dominated by ions [38].

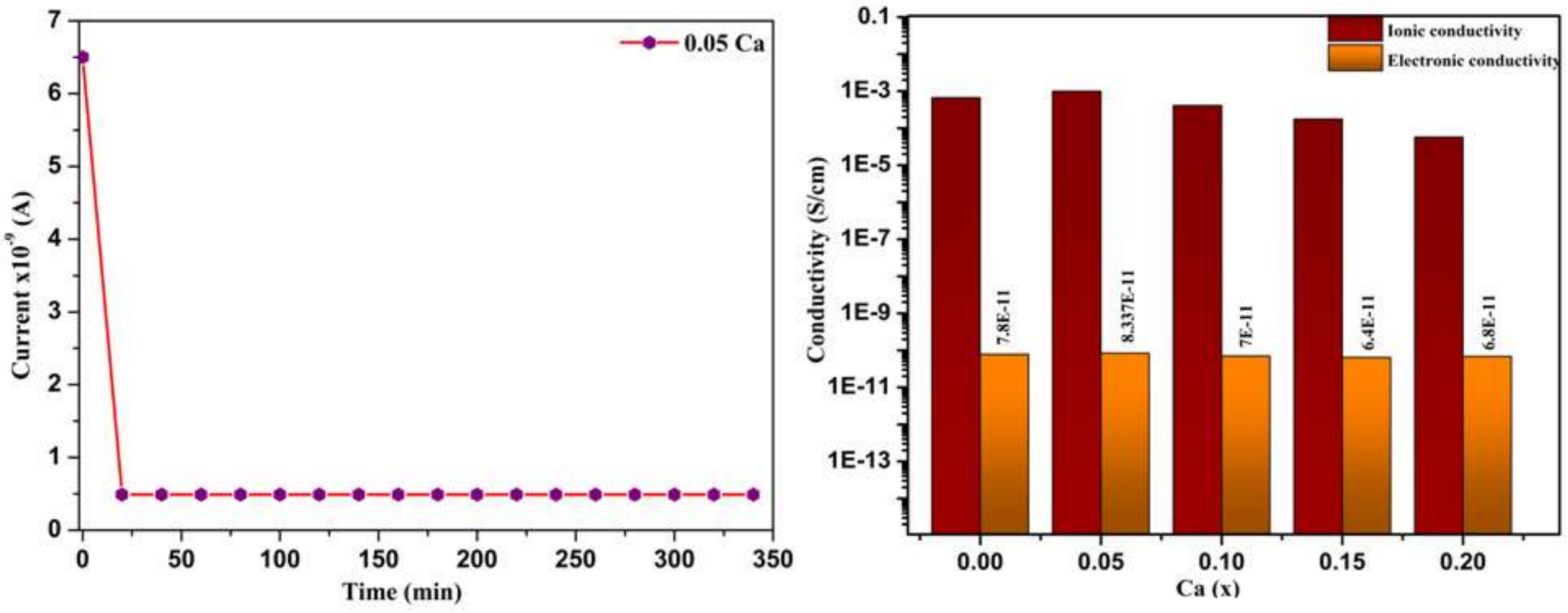

**(a)** **(b)**

**Fig. 7: (a) DC conductivity plot of 0.05 Ca ceramic sample. (b) Comparison between ionic and electronic conductivities of all the samples of $Li_{6.55+x}Ge_{0.05}La_{3-x}Ca_xZr_{1.75}Ta_{0.25}O_{12}$ with x = 0 – 0.20.**

## 4. Conclusions

All the prepared ceramic samples in the series $Li_{6.55+x}Ge_{0.05}La_{3-x}Ca_xZr_{1.75}Ta_{0.25}O_{12}$ (x = 0– 0.20), synthesized using the conventional solid state reaction method exhibit the cubic structure. Among all the ceramic samples, 0.05 Ca containing ceramic sample shows the dense microstructure with the highest relative density of 95.09%. This has been attributed to the optimum Ca incorporation that helps in the densification of the structure and prevents the dendrite growth formation. The highest density is favorable for Li-ion conduction. The highest conductivity of 9.95 x $10^{-4}$ S/cm at room temperature is achieved for 0.05 Ca ceramic. It is supported by the highest relative density. The highest ionic conductivity is associated with the minimum activation energy of 0.23 eV. This has been attributed to the insertion of $Ca^{2+}$ at the $La^{3+}$ site. It leads to the increase in the Li content with the decrease in the energy barrier for Li-ion. The DC polarization measurements confirmed the ionic nature of conductivity in 0.05 Ca ceramic sample. Thus, 0.05 Ca ceramic is a promising candidate as solid electrolyte for ASSLIBs.

**Acknowledgment**

One of the authors likes to acknowledge VNIT, Nagpur, for giving a Ph.D. fellowship. The author is thankful to the Department of Physics at VNIT, Nagpur, for providing the XRD facility, and for the support of DST FIST project number SR/FST/PSI/2017/5(C).

**Conflict of Interest**

Authors declare that there are no conflicts of interest.

**Funding**

This research did not receive any specific grant from funding agencies in the public, commercial, or not-for-profit sectors.

**References**

1. Larcher D, Tarascon JM. Towards greener and more sustainable batteries for electrical energy storage. *Nat Chem*. 2015;7(1):19–29. https://doi.org/10.1038/nchem.2085

2. Aykol M, Herring P, Anapolsky A. Machine learning for continuous innovation in battery technologies. *Nat Rev Mater*. 2020;5(10):725–727. https://doi.org/10.1038/s41578-020-0216-y

3. Luo Y, Zhang Q, Shen A, Shen M, Xie D, Yan Y. Calcium-doping effects on structure and electric performances of garnet-type Li6.6La3Zr1.6Sb0.4O12 solid-state electrolytes. *Solid State Ionics*. 2022;374(2):4–11. https://doi.org/10.1016/j.ssi.2021.115812

4. Wang Y, Richards WD, Ong SP, *et al.* Design principles for solid-state lithium superionic

conductors. *Nat Mater*. 2015;14(10):1026–1031. https://doi.org/10.1038/nmat4369

5. Li J, Cheng Y, Ai L, *et al.* 3D simulation on the internal distributed properties of lithium-ion battery with planar tabbed configuration. *J Power Sources*. 2015;293:993–1005. https://doi.org/10.1016/j.jpowsour.2015.06.034

6. Aote M, Deshpande A V. Enhancement in Li-ion conductivity through Co-doping of Ge and Ta in garnet Li7La3Zr2O12 solid electrolyte. *Ceram Int*. 2023;(August). https://doi.org/10.1016/j.ceramint.2023.09.330

7. Zhang Y, Deng J, Hu D, *et al.* Synergistic regulation of garnet-type Ta-doped Li7La3Zr2O12 solid electrolyte by Li+ concentration and Li+ transport channel size. *Electrochim Acta*. 2019;296:823–829. https://doi.org/10.1016/j.electacta.2018.11.136

8. Chen YT, Jena A, Pang WK, *et al.* Voltammetric Enhancement of Li-Ion Conduction in Al-Doped Li7-xLa3Zr2O12 Solid Electrolyte. *J Phys Chem C*. 2017;121(29):15565–15573. https://doi.org/10.1021/acs.jpcc.7b04004

9. Wakudkar P, Deshpande A V. Effect of Li4SiO4 addition in Li6.22Al0.16La3Zr1.7Ta0.3O12 garnet type solid electrolyte for lithium ion battery application. *Ceram Int*. 2019;45(16):20113–20120. https://doi.org/10.1016/j.ceramint.2019.06.276

10. Society CC. 石榴石型固态电解质 Li 7 La 3 Zr 2 O 12 的改性研究 现状. 2022;1861–1870.

11. Zhuang L, Huang X, Lu Y, *et al.* Phase transformation and grain-boundary segregation in Al-Doped Li7La3Zr2O12 ceramics. *Ceram Int*. 2021;47(16):22768–22775. https://doi.org/10.1016/j.ceramint.2021.04.295

12. Chen C, Sun Y, He L, *et al.* Microstructural and Electrochemical Properties of Al- And Ga-Doped Li7La3Zr2O12 Garnet Solid Electrolytes. *ACS Appl Energy Mater*. 2020;3(5):4708–4719. https://doi.org/10.1021/acsaem.0c00347

13. Brugge RH, Kilner JA, Aguadero A. Germanium as a donor dopant in garnet electrolytes. *Solid State Ionics*. 2019;337(April):154–160. https://doi.org/10.1016/j.ssi.2019.04.021

14. Liu X, Gao M, Liu Y, Xiong L, Chen J. Improving the room temperature ionic conductivity of Al-Li 7 La 3 Zr 2 O 12 ceramics by Ba and Y or Ba and W co-doping. *Ceram Int*. 2019;45(10):13488–13495. https://doi.org/10.1016/j.ceramint.2019.04.052

15. Dubey BP, Sahoo A, Thangadurai V, Sharma Y. Morphological, dielectric and transport properties of garnet-type Li6.25+yAl0.25La3Zr2-yMnyO12 (y = 0, 0.05, 0.1, and 0.2). *Solid State Ionics*. 2020;351(January):115339. https://doi.org/10.1016/j.ssi.2020.115339

16. Liang X, Li X, Wu X, Gai Q, Zhang X. Effect of Zn-Incorporation into Li7La3Zr2O12 Solid Electrolyte on Structural Stability and Electrical Conductivity: First-Principles Study. *Int J Electrochem Sci*. 2020;15(11):11123–11136. https://doi.org/10.20964/2020.11.38

17. Dermenci KB, Buluç AF, Turan S. The effect of limonite addition on the performance of Li7La3Zr2O12. *Ceram Int*. 2019;45(17):21401–21408. https://doi.org/10.1016/j.ceramint.2019.07.128

18. Li Y, Wang Z, Cao Y, *et al.* W-Doped Li7La3Zr2O12 Ceramic Electrolytes for Solid State Li-ion Batteries. *Electrochim Acta*. 2015;180:37–42. https://doi.org/10.1016/j.electacta.2015.08.046

19. Luo Y, Li X, Zhang Y, Ge L, Chen H, Guo L. Electrochemical properties and structural stability of Ga- and Y- co-doping in Li7La3Zr2O12 ceramic electrolytes for lithium-ion batteries. *Electrochim Acta*. 2019;294:217–225. https://doi.org/10.1016/j.electacta.2018.10.078

20. Lin C, Tang Y, Song J, Han L, Yu J, Lu A. Sr- and Nb-co-doped Li7La3Zr2O12 solid electrolyte with Al2O3 addition towards high ionic conductivity. *Appl Phys A Mater Sci Process*. 2018;124(6):1–8. https://doi.org/10.1007/s00339-018-1849-1

21. Miara LJ, Ong SP, Mo Y, *et al.* Effect of Rb and Ta doping on the ionic conductivity and stability of the garnet Li7+2x-y(La3-xRbx)(Zr 2-yTay)O12 ($0 \le x \le 0.375$, $0 \le y \le 1$) Superionic conductor: A first principles investigation. *Chem Mater*. 2013;25(15):3048–3055. https://doi.org/10.1021/cm401232r

22. Chen X, Cao T, Xue M, Lv H, Li B, Zhang C. Improved room temperature ionic conductivity of Ta and Ca doped Li7La3Zr2O12 via a modified solution method. *Solid State Ionics*. 2018;314(November 2017):92–97. https://doi.org/10.1016/j.ssi.2017.11.027

23. Hosokawa H, Takeda A, Inada R, Sakurai Y. Tolerance for Li dendrite penetration in Ta-doped Li7La3Zr2O12 solid electrolytes sintered with Li2.3C0.7B0.3O3 additive. *Mater Lett*. 2020;279:128481. https://doi.org/10.1016/j.matlet.2020.128481

24. Liu XZ, Ding L, Liu YZ, Xiong LP, Chen J, Luo XL. Room-temperature ionic conductivity of Ba, Y, Al co-doped Li7La3Zr2O12 solid electrolyte after sintering. *Rare Met*. 2021;40(8):2301–2306. https://doi.org/10.1007/s12598-020-01526-x

25. Meesala Y, Liao YK, Jena A, *et al.* An efficient multi-doping strategy to enhance Li-ion conductivity in the garnet-type solid electrolyte Li7La3Zr2O12. *J Mater Chem A*. 2019;7(14):8589–8601. https://doi.org/10.1039/c9ta00417c

26. Russo I. Garnet Ceramic Electrolytes for Lithium Ion Batteries. 2014;1–28.

27. Qian D, Xu B, Cho HM, Hatsukade T, Carroll KJ, Meng YS. Lithium lanthanum titanium oxides: A fast ionic conductive coating for lithium-ion battery cathodes. *Chem Mater*. 2012;24(14):2744–2751. https://doi.org/10.1021/cm300929r

28. Wang C, Lin PP, Gong Y, Liu ZG, Lin TS, He P. Synergistic impacts of Ca2+ and Ta5+ dopants on electrical performance of garnet-type electrolytes. *J Alloys Compd*. 2021;879. https://doi.org/10.1016/j.jallcom.2021.160420

29. Chen X, Wang T, Lu W, *et al.* Synthesis of Ta and Ca doped Li7La3Zr2O12 solid-state electrolyte via simple solution method and its application in suppressing shuttle effect of Li-S battery. *J Alloys Compd*. 2018;744:386–394. https://doi.org/10.1016/j.jallcom.2018.02.134

30. Song S, Sheptyakov D, Korsunsky AM, Duong HM, Lu L. High Li ion conductivity in a garnet-type solid electrolyte via unusual site occupation of the doping Ca ions. *Mater Des*. 2016;93:232–237. https://doi.org/10.1016/j.matdes.2015.12.149

31. Abrha LH, Hagos TT, Nikodimos Y, *et al.* Dual-Doped Cubic Garnet Solid Electrolytes with Superior Air Stability. *ACS Appl Mater Interfaces*. 2020;12(23):25709–25717. https://doi.org/10.1021/acsami.0c01289

32. El-Shinawi H, Cussen EJ, Corr SA. Enhancement of the lithium ion conductivity of Ta-doped Li7La3Zr2O12 by incorporation of calcium. *Dalt Trans*. 2017;46(29):9415–9419. https://doi.org/10.1039/c7dt01573a

33. Ning T, Zhang Y, Zhang Q, *et al.* The effect of a Ta, Sr co-doping strategy on physical and electrochemical properties of Li7La3Zr2O12 electrolytes. *Solid State Ionics*. 2022;379(March):115917. https://doi.org/10.1016/j.ssi.2022.115917

34. Huang X, Song Z, Xiu T, Badding ME, Wen Z. Sintering, micro-structure and Li+ conductivity of Li7−xLa3Zr2−xNbxO12/MgO (x = 0.2–0.7) Li-Garnet composite ceramics. *Ceram Int*. 2019;45(1):56–63. https://doi.org/10.1016/j.ceramint.2018.09.133

35. Khanmirzaei MH, Ramesh S, Ramesh K. Polymer electrolyte based dye-sensitized solar cell with rice starch and 1-methyl-3-propylimidazolium iodide ionic liquid. *Mater Des*. 2015;85:833–837. https://doi.org/10.1016/j.matdes.2015.06.113

36. Song S, Lu J, Zheng F, Duong HM, Lu L. A facile strategy to achieve high conduction and excellent chemical stability of lithium solid electrolytes. *RSC Adv*. 2015;5(9):6588–6594. https://doi.org/10.1039/c4ra11287c

37. Zhu Y, Wu S, Pan Y, Zhang X, Yan Z, Xiang Y. Reduced Energy Barrier for Li+ Transport Across Grain Boundaries with Amorphous Domains in LLZO Thin Films. *Nanoscale Res Lett*. 2020;15(1). https://doi.org/10.1186/s11671-020-03378-x

38. Enkhbayar E, Kim J. Study of Codoping Effects of Ta5+and Ga3+on Garnet Li7La3Zr2O12. *ACS Omega*. 2022;7(50):47265–47273. https://doi.org/10.1021/acsomega.2c06544